\documentclass[manuscript]{aastex}
\usepackage{natbib}
\usepackage{booktabs}
\usepackage{threeparttable}
\usepackage{amsmath,bm}

\begin{document}

\title{Magnetic helicity signature and its role in regulating magnetic energy spectra and proton temperatures in the solar wind}

\author{G. Q. Zhao\altaffilmark{1,2}, Y. Lin\altaffilmark{3}, X. Y. Wang\altaffilmark{3}, H. Q. Feng\altaffilmark{1}, D. J. Wu\altaffilmark{4}, H. B. Li\altaffilmark{1}, A. Zhao\altaffilmark{1}, and Q. Liu\altaffilmark{1}}

\affil{$^1$Institute of Space Physics, Luoyang Normal University, Luoyang, China}
\affil{$^2$Henan Key Laboratory of Electromagnetic Transformation and Detection, Luoyang, China}
\affil{$^3$Physics Department, Auburn University, Auburn, USA}
\affil{$^4$Purple Mountain Observatory, CAS, Nanjing, China}
\altaffiltext{1}{Correspondence should be sent to: zgqisp@163.com}

\begin{abstract}
In a previous paper, we found that proton temperatures are clearly associated with the proton-scale turbulence in the solar wind, and magnetic helicity signature appears to be an important indicator in the association. Based on 15 years of in situ measurements, the present paper further investigates the magnetic helicity of solar wind turbulence and its role in regulating magnetic energy spectra and proton temperatures. Results show that the presence of the helicity signature is very common in solar wind turbulence at scales $0.3 \lesssim  k\rho_p \lesssim 1$, with $k$ being the wavenumber and $\rho_p$ the proton gyroradius. The sign of the helicity is mostly positive, indicating the dominance of right-handed polarization of the turbulence. The helicity magnitude usually increases with $k$ and $\beta_{{\parallel}p}$ (the proton parallel beta) when $k\rho_p$ and $\beta_{{\parallel}p}$ are less than unity. As helicity magnitude increases, the power index of the energy spectrum becomes more negative, and the proton temperatures $T_{{\perp}p}$ and $T_{{\parallel}p}$ rise significantly, where $T_{{\perp}p}$ and $T_{{\parallel}p}$ are the perpendicular and parallel temperatures with respect to the background magnetic field. In particular, the rise of $T_{{\perp}p}$ is faster than $T_{{\parallel}p}$ when $\beta_{{\parallel}p} < 1$ is satisfied. The faster rise of $T_{{\perp}p}$ with the helicity magnitude may be interpreted as the result of the preferentially perpendicular heating of solar wind protons by kinetic Alfv\'en wave (KAW) turbulence.
\end{abstract}


\section{Introduction}
It has long been known that the solar wind undergos nonadiabatic expansion, with proton temperatures usually much higher than theoretical prediction \citep[e.g.,][]{mar82p52,gaz82p43}.
Early researches revealed that the average dependence of proton temperatures on the heliocentric radial distance follows a power law $r^{n_T}$, where $n_T \gtrsim -1$ is satisfied for the radial distance between 0.3 and 9 au. This index is greater than the index expected for isotropic adiabatic expansion, i.e., $-4/3$. Moreover, considering the weakly collisional solar wind with proton distributions far from thermodynamic equilibrium, the double-adiabatic theory predicts the perpendicular proton core temperature decreases as $r^{-2}$ \citep{che56p12,mat12p73}, whereas observations showed that this temperature decreases significantly more slowly. An index around $-0.9$ was often reported \citep{hel11p05,mat13p71,per19p30,hua20p70}. This behavior implies that some heating process must be at work in the solar wind.

Many heating sources have been proposed to explain the nonadiabatic behavior of the expansion. They are ion cyclotron waves evolved from Alfv\'en waves \citep{mar82p30,tuc97p63,hol02p47}, electron heat flux generated by small-scale reconnections \citep{mar02p43,mar04p12}, drift ion cyclotron modes on account of density structures/gradient \citep{vra08p53,vra09p18}, fast magnetosonic waves produced by ion beams \citep{luq06p01,hel11p01}, inertial-range intermittency \citep{osm11p11,osm12p02}, or small-scale turbulence \citep{cha09p68,kiy15p55}. Among them much attention has been paid to ion cyclotron waves and small-scale turbulence in recent years \citep[][]{kas13p02,cra14p16,oza15p77,kiy15p55,mat16p07,hug17p14,ise19p63,arz19p53,hej19p21,hej20p43,che20p53,bow20p66,hua20p03,zha20p14}.

The small-scale turbulence is of interest in the present paper. The solar wind magnetic fluctuations are inherently turbulent. Large-scale fluctuations will cascade within the inertial range, and become proton-scale turbulence where turbulent dissipation and heating would be expected \citep{ale13p01,bru13p02,ver19p05}. Existing researches have shown that the energy transfer by the cascade in the inertial range is efficient to account for the proton temperature radial profile of the solar wind \citep{mac08p44,sta09p19,ban20p48}. As for the turbulent dissipation, it is unclear what is the specific mechanism to convert the turbulent energy into particle kinetic energy. Several mechanisms including cyclotron damping \citep{smi12p08,cra14p16,woo18p49}, Landau damping \citep{lea99p31,hej15p76,how18p05}, non-resonant stochastic heating \citep[][]{joh01p21,cha10p03,mar19p43}, and plasma coherent structures including magnetic vortices, reconnecting current sheets, and shocks \citep{bru03p30,per12p01,wan19p22} were invoked.

In a recent study, based on 11 years of in situ measurements, we presented evidence that the solar wind is heated by the proton-scale turbulence \citep{zha20}. It was shown that the proton perpendicular temperature is clearly associated with the proton-scale turbulence. A positive power-law correlation between the perpendicular temperature and turbulent magnetic energy at proton scales was found, and a scenario for the turbulence and heating was proposed. On the other hand, our results imply that magnetic helicity tends to play an important role in indicating the heating. The magnetic helicity is a measure of the spatial handedness
of the fluctuating magnetic field \citep{mat82p11}. Statistically, the majority of nearly collisionless solar wind turbulence is characterized by magnetic helicity signature. Enhanced helicity signature appears to result in steeper magnetic energy spectra at proton scales and favor a better correlation of the temperature with the magnetic energy. Despite these findings, there is still room to exploit the magnetic helicity. The details for the helicity distribution, and for the dependence of the energy spectra on the helicity, however, are absent. The dependence of proton temperatures (perpendicular and parallel components) on the helicity is also not discussed in \citet{zha20}.

Based on in situ measurements over 15 years, the present paper aims to further exploit the magnetic helicity and show its role in regulating magnetic energy spectra and proton temperatures in the solar wind. The paper is organized as follows. The data and analysis methods used in this paper are introduced in Section 2. Statistical results concerning the magnetic helicity, magnetic energy spectra, and proton temperatures are presented in Section 3. Section 4 is the summary and discussion.

\section{Data and Analysis Methods}
The data used in the present paper are over a long time period from 2004 June to 2019 May. They are from the \emph{Wind} spacecraft, a comprehensive solar wind laboratory in a halo orbit around the L1 Lagrange point. The magnetic fields are sampled at a cadence of 0.092 s \citep{lep95p07}, and the plasma data are at a cadence of 92 s \citep{ogi95p55}. The proton temperatures are yielded through a nonlinear-least-squares bi-Maxwellian fit of ion spectrum from the Faraday cup \citep{kas06p05}.
The survey is through dividing the long time series into consecutive and overlapping time segments. Each time segment has a span of 200 s, and the overlap time is set to be 100 s.
In each segment with data available, the magnetic energy spectrum is produced by standard fast Fourier transform technique. The plasma parameters are obtained as average values over the time segment. They are composed of the proton density $N_p$, perpendicular and parallel thermal velocities $w_{{\perp}p}$ and $w_{{\parallel}p}$, and bulk velocity ${\bm V}_p$, where $\perp$ and $\parallel$ are with respect to the background magnetic field ${\bm B}_0$. Following the paper \citep{zha20}, segments with $A_c < 0.1$ are selected to restrict the study to the solar wind with negligible collision effects, where $A_c$ is the Coulomb collisional age \citep{liv86p45}. The angle between ${\bm B}_0$ and ${\bm V}_p$ is also required in the range from $60^\circ$ to $120^\circ$, which could reduce the possible heating/cooling effects due to the alpha$-$proton differential flow \citep{zha19p60,zha20p14}. In total about $3.7 \times 10^5$ time segments satisfying these constraints are selected.

Magnetic helicity has been widely used to indicate the presence of circularly/elliptically polarized waves in the solar wind \citep{hej11p85,zha18p15,zha19p75,woo19p53}. The helicity is reduced for the magnetic field measured by a single spacecraft. It can be expressed as $k{H}_m^r({k})/P(k)$, where $k$ is the reduced wavenumber, and ${H}_m^r$ and $P$ are the reduced fluctuating magnetic helicity and the magnetic energy at wavenumber $k$, respectively \citep{mat82p11}. It can be further expressed as a function of frequency associated with the spacecraft time series when Taylor frozen-in-flow hypothesis holds, where the frequency is related to wavenumber \citep{tay38p76,hej11p85}. The equation for the magnetic helicity used in this paper is finally written as \citep{mat82p11,zha20}
\begin{eqnarray}
\sigma_f=\frac{2\mathrm{Im}[B_y(f){\cdot}B_z^*(f)]}{\sum\limits_{i=j}[B_i(f){\cdot}B_j^*(f)]},
\label{sigma}
\end{eqnarray}
where $\mathrm{Im}$ means the imaginary part, $B_i{\cdot}B_j^*$ are the elements of the energy spectral tensor coming from Fourier spectra of time series of magnetic fields, and the subscripts $i$ and $j$ indicate the components of the magnetic field vector in the GSE coordinate system.
Equation \eqref{sigma} also means that the helicity is normalized, with values in the range from $-1$ to 1. Here a negative (positive) value implies left(right)-handed polarization with respective to the $x$ direction that points to the Sun. In order to obtain the polarization sense with respective to the background magnetic field, the helicity will be multiplied by $-1$ when the background field points outward from the Sun. After this step the negative (positive) helicity will correspond left(right)-handed polarization with respective to the background field.

\begin{figure}
\epsscale{0.6} \plotone{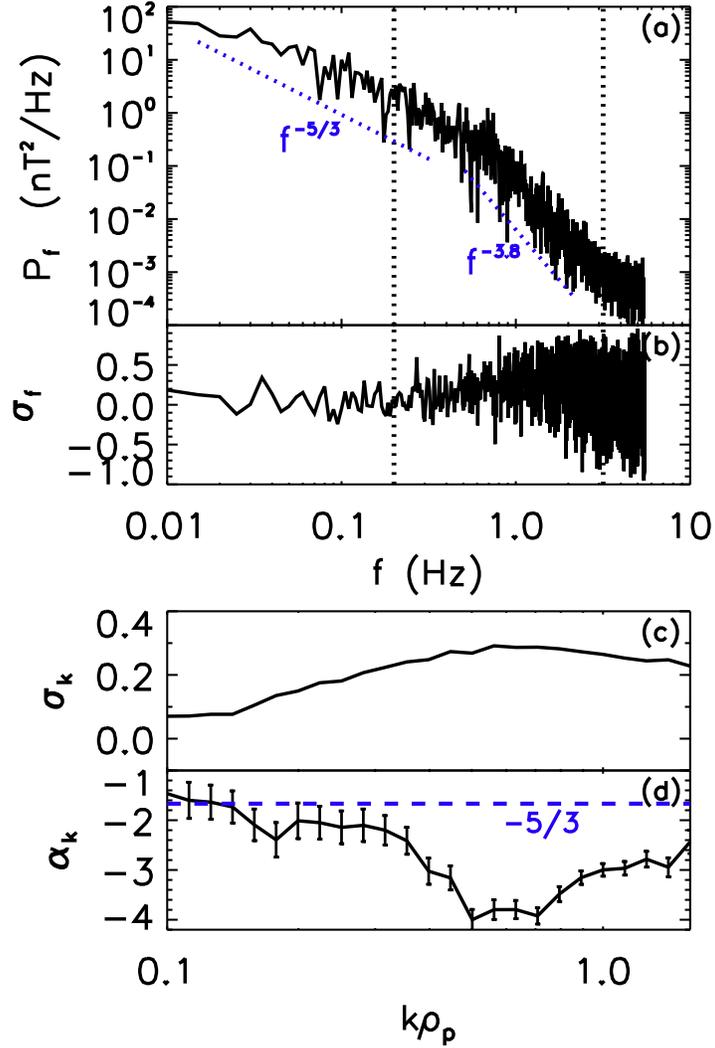} \caption{An example to illustrate (a) magnetic energy spectrum  $P_f$ in frequency domain, (b) magnetic helicity $\sigma_f$ in frequency domain, (c) local average magnetic helicity $\sigma_k$ in wavenumber domain, (d) local magnetic spectral index $\alpha_k$ in wavenumber domain. Two vertical dotted lines in panels (a) and (b) indicate the range plotted in panels (c) and (d).}
\end{figure}

Using the data on 23 June 2005, 19:04:40$-$19:08:00 UT (one time segment), Figure 1 is presented to illustrate the helicity as well as the spectral index of the magnetic fluctuation. Panels (a) and (b) plot the energy spectrum ($P_f$) and the helicity ($\sigma_f$) of the fluctuation in the frequency domain. The energy spectrum is characterize by two power laws. One is with an index nearly $-5/3$ in the lower frequency regime while the other is with an index about $-3.8$ when the frequency exceeds 0.7 Hz. (The spectrum flattens significantly when the frequency exceeds 3 Hz, which is probably due to the instrument noise and/or aliasing.) As the energy spectrum steepens, the helicity has a trend to rise considerably. The two vertical dotted lines in panel (a) and (b) indicate the range of our interest. Same to the paper \citep{zha20}, the left line indicates the wavenumber $k\rho_p = 0.1$, where $\rho_p=w_{{\perp}p}/\Omega_{p}$ is the proton thermal gyroradius and $\Omega_{p}$ is the proton cyclotron frequency, and the right line is chosen accordingly, with the spectral energy $P_f > 10^{-3}$ nT$^2$/Hz so that signal level is much higher than the instrument noise level \citep{lep95p07}. The conversion from the frequency domain to the wavenumber domain is conducted according to the Taylor frozen-in-flow hypothesis, ${2\pi{f}}=kV_p$ \citep{tay38p76}. This step should be meaningful for an analysis of various observations with different plasma parameters. Then the helicity ($\sigma_k$) in the wavenumber domain can be presented, where an averaging operation over $fe^{-0.5} \leq f \leq fe^{0.5}$ is conducted to produce a smoothed helicity spectrum, as shown in panel (c). Panel (d) displays the local spectral index ($\alpha_k$) that is yielded by fitting the energy spectrum over the same frequency range for the averaging. The blue horizontal dashed line in this panel marks the constant $-5/3$. Comparing panel (d) with panel (c), one may note that the more negative $\alpha_k$ occurs with the larger $\sigma_k$.

\section{Statistical Results}
Based on the data set described in Section 2, statistical investigations on the magnetic helicity, magnetic energy spectra, and proton temperatures are conducted.  In the investigations the sign of the helicity is taken into account, which was ignored in our previous study \citep{zha20}.
Subsection 3.1 presents the statistical results for magnetic helicity distributions. Subsections 3.2 and 3.3 display the results for magnetic spectral indices and proton temperatures regulated by the magnetic helicity, respectively.

\subsection{Distributions of Magnetic Helicity $\sigma_k$}
The distributions of $\sigma_k$ depend on the wavenumber and the proton beta. Figure 2 plots the distributions at a fixed wavenumber $k\rho_p=0.8$, but for different beta ranges, where the bin of 0.02 for $\sigma_k$ has been used. Panel (a) in Figure 2 is for all $\beta_{{\parallel}p}$ observed, while panels (b) and (c) are for the data subsets with $\beta_{{\parallel}p} < 0.3$ and $0.8 < \beta_{{\parallel}p} < 1$, respectively, where $\beta_{{\parallel}p}=w_{{\parallel}p}^2/v_A^2$ is the proton parallel beta. One may first find that the distributions are asymmetrical with respect to $\sigma_k=0$, since the data with positive $\sigma_k$ are much more than those with negative $\sigma_k$.
There are two peaks arising at $\sigma_k \simeq \pm 0.22$, although they are weak in the case of lower beta (pane (b)). This result implies that the majority of the data are characterized by a considerable helicity with the magnitude greater than 0.1. On the other hand, the data size in a bin drops dramatically when the helicity magnitude approaches to 0.4, which happens nearly irrespective of $\beta_{{\parallel}p}$. 

\begin{figure}
\epsscale{1.0} \plotone{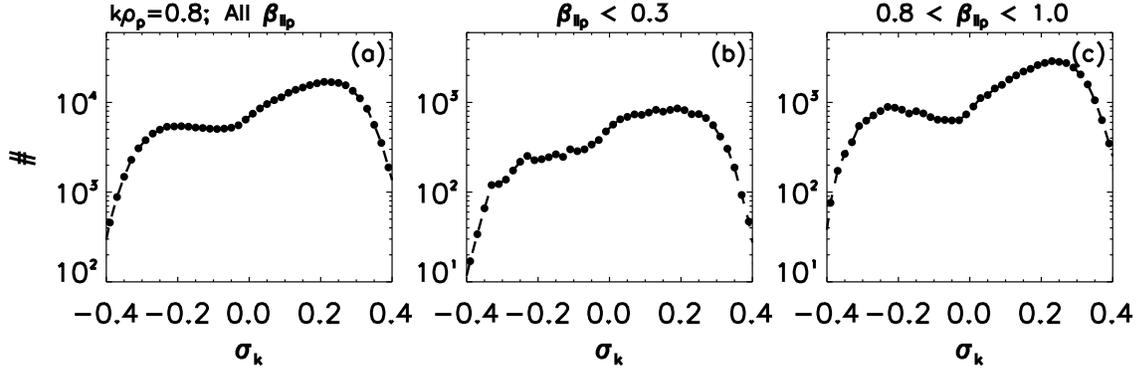} \caption{Distributions of $\sigma_k$ at $k\rho_p=0.8$ for cases of (a) all $\beta_{{\parallel}p}$, (b) $\beta_{{\parallel}p} < 0.3$, (c) $0.8 < \beta_{{\parallel}p} < 1$.}
\end{figure}

\begin{figure}
\epsscale{1.0} \plotone{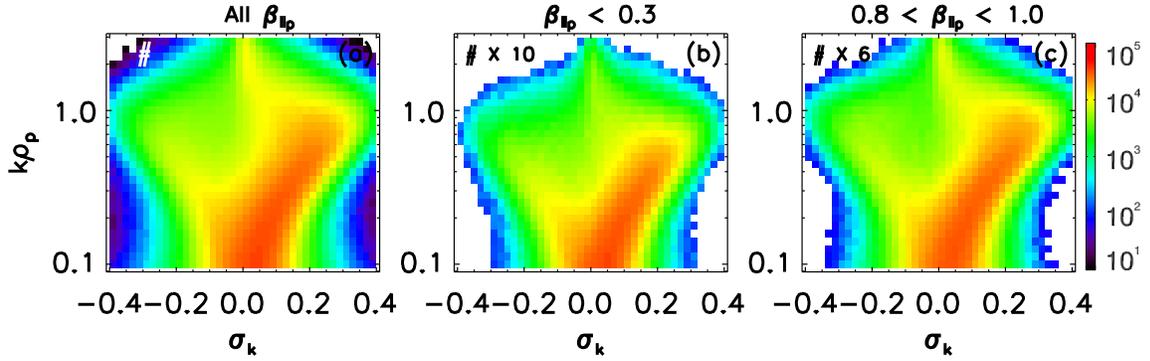} \caption{Color scale plot of distributions of $\sigma_k$ at various $k$ for cases of (a) all $\beta_{{\parallel}p}$, (b) $\beta_{{\parallel}p} < 0.3$, (c) $0.8 < \beta_{{\parallel}p} < 1$.}
\end{figure}

\begin{figure}
\epsscale{0.6} \plotone{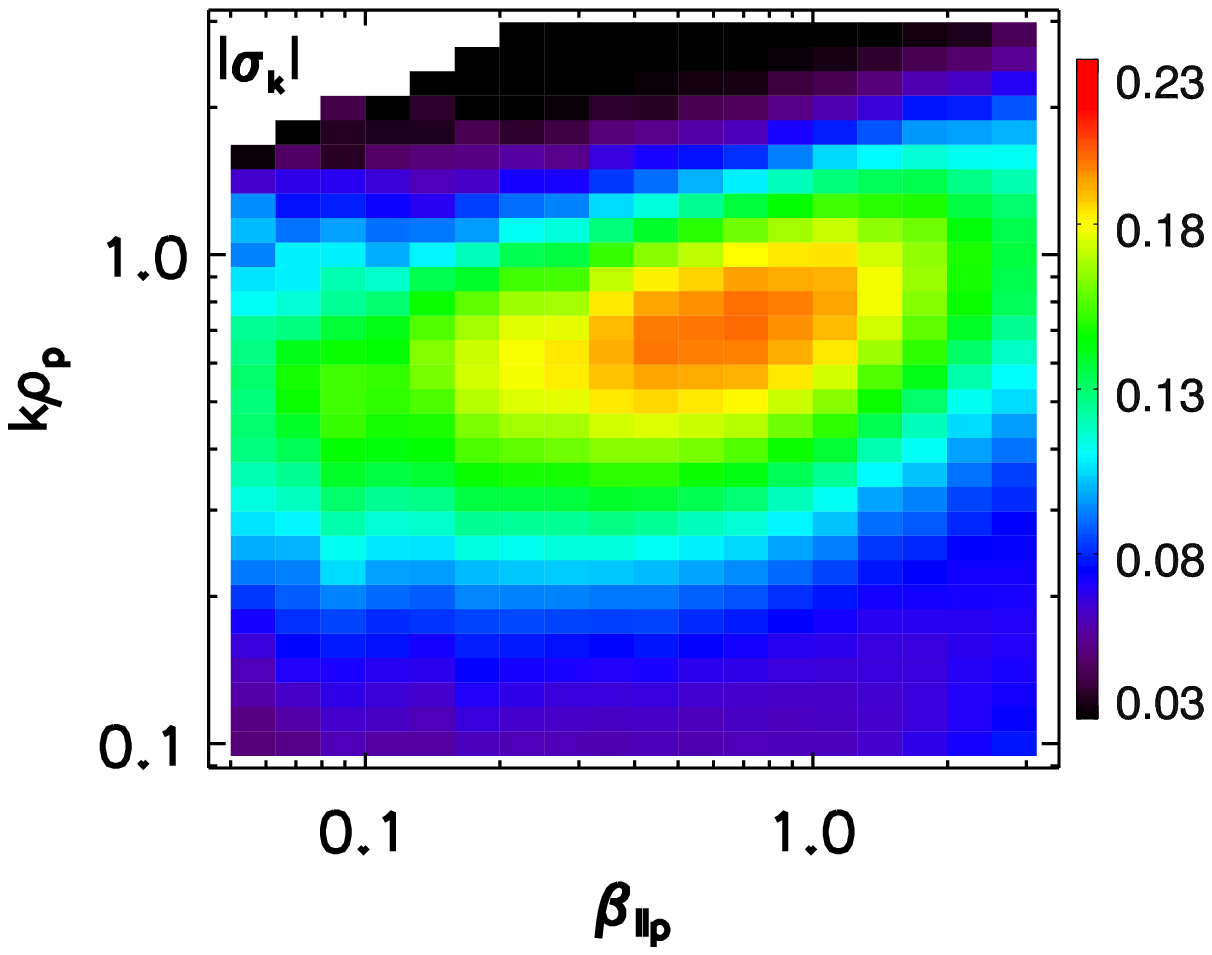} \caption{Color scale plot of medians of $|\sigma_k|$ in the ($\beta_{{\parallel}p}$, $k$) space.}
\end{figure}

Figure 3 displays the color plot of distributions of $\sigma_k$ for various $k$. Three panels correspond to three $\beta_{{\parallel}p}$ ranges as those in Figure 2, and the red color means the peak of the distribution. In order to share a common color bar, the data numbers in panels (b) and (c) have been amplified by 10 times and 6 times, respectively.
One can see that the helicity distribution significantly depends on the wavenumber in each panel. For a small wavenumber $k\rho_p=0.1$, the peak of the distribution occurs at $\sigma_k \sim 0$, while two peaks with finite $\sigma_k$ arise when $k\rho_p \gtrsim 0.3$. One can also see that the peak with positive $\sigma_k$ dominates the distribution and the other peak is much weak. As $k\rho_p$ increases up to 0.7$\sim$1 (depending on $\beta_{{\parallel}p}$), the magnitude of $\sigma_k$ begins to decrease rapidly. The cause of the decrease might be complex, possibly due to the instrument noise, aliasing, and/or great balance of wave turbulence \citep{hej12p86,mar13p62,kle14p38}. In addition, the data number decreases significantly when $k\rho_p$ exceeds some value, 0.7$\sim$1. This is because a lot of time segments described in Section 2 have spectral energies lower than the threshold $10^{-3}$ nT$^2$/Hz at the larger $k$, which are discarded in the statistics to reduce the effect of noise on the results.

To explore the role of $\beta_{{\parallel}p}$ in determining the helicity, Figure 4 presents medians of $|\sigma_k|$ with respect to ($\beta_{{\parallel}p}$, $k$). Here only the magnitude of $\sigma_k$ is used, since the sign of $\sigma_k$ just means the polarization sense.
Result shows that statistically $|\sigma_k|$ increases with $k$ when $k\rho_p \lesssim 1$ and decreases if $k\rho_p > 1$, which is consistent with Figure 3. For a given $k$ with $0.3 \lesssim k\rho_p \lesssim 1$,  $\sigma_k$ increases with $\beta_{{\parallel}p}$ when $\beta_{{\parallel}p} \lesssim 1$ is satisfied. For $\beta_{{\parallel}p} > 1$, $\sigma_k$ decreases considerably. The increase of $|\sigma_k|$ with $\beta_{{\parallel}p}$ is in agreement with the results obtained by \citet{mar13p62,mar16p015}, who calculated the magnetic helicity via hybrid numerical simulations of two-dimensional turbulence for three beta values, i.e., 0.15, 0.5, and 0.65. 

\subsection{Spectral Indices Regulated by $\sigma_k$}
Existing literatures show that spectral indices of the proton-scale magnetic fluctuations in the solar wind take vales usually between $-2$ and $-4$ \citep{smi06p85,sah13p15,bru14p15,pig20p84}.
Our statistics is in agreement with previous results. In particular, it is found that the spectral indices can be well regulated by $\sigma_k$, especially in the case of low beta. Figure 5 displays the distributions of ($\sigma_k$, $\alpha_k$), where $\alpha_k$ is the spectral index. In the figure panels (a)$-$(c) correspond to three cases of all $\beta_{{\parallel}p}$, $\beta_{{\parallel}p} < 0.3$, and $0.8 < \beta_{{\parallel}p} < 1$, respectively, and the wavenumber $k\rho_p=1$ has been fixed. One can see that $\alpha_k$ decreases with $|\sigma_k|$ overall, which implies that a larger $|\sigma_k|$ account for a steeper energy spectrum. We fit the data by $\alpha_k=a\sigma_k + b$, distinguished between $\sigma_k > 0$ and $\sigma_k < 0$. The fitted parameters are presented in Table 1. It shows that $\alpha_k$, relative to the situation for $\sigma_k < 0$, tends to have stronger dependence on $\sigma_k$ when $\sigma_k > 0$. It also appears that the dependence is stronger if $\beta_{{\parallel}p}$ is lower, and
the strongest dependence of $\alpha_k$ on $\sigma_k$ occurs in the case of $\beta_{{\parallel}p} < 0.3$, with $\alpha_k=(-3.5\pm0.04)\sigma_k -2.1\pm0.01$ for $\sigma_k > 0$.

\begin{figure}
\epsscale{1.} \plotone{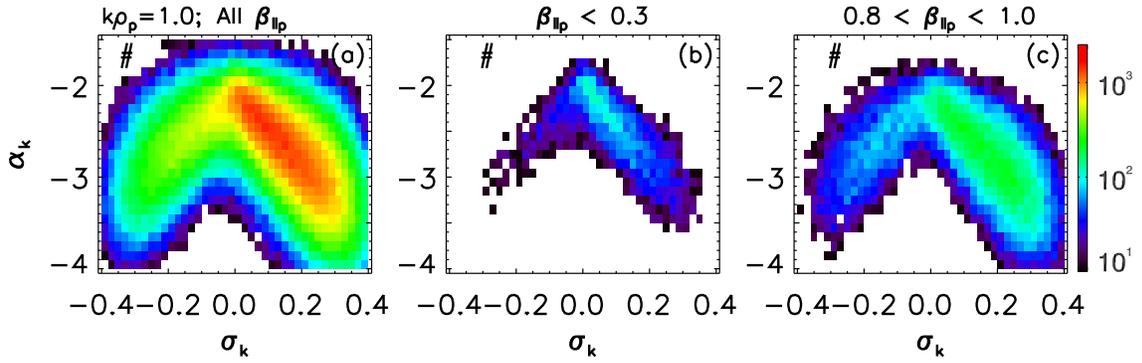} \caption{Distributions of ($\sigma_k$, $\alpha_k$) at $k\rho_p=1$ for cases of (a) all $\beta_{{\parallel}p}$, (b) $\beta_{{\parallel}p} < 0.3$, (c) $0.8 < \beta_{{\parallel}p} < 1$.}
\end{figure}

\begin{table}
\caption{Fitted parameters for the expression of $\alpha_k=a\sigma_k + b$ at $k\rho_p=1$.}
\centering
\begin{tabular}{l c c c c}
\hline\hline
& \multicolumn{2}{c}{$\sigma_k > 0$} & \multicolumn{2}{c}{$\sigma_k < 0$} \\ [-4.0ex]
Case \\ [-0.5ex]
\cline{2-5} \\ [-5.0ex]
& $a$ & $b$ & $a$ & $b$ \\ [-0.5ex]
\hline
All $\beta_{{\parallel}p}$   & $-2.9\pm0.009$ & $-2.2\pm0.002$ & $2.6\pm0.015$ & $-2.2\pm0.003$ \\  [-1.0ex]
$\beta_{{\parallel}p} < 0.3$ & $-3.5\pm0.039$ & $-2.1\pm0.007$ & $3.2\pm0.062$ & $-2.1\pm0.010$ \\ [-1.0ex]
$0.8 < \beta_{{\parallel}p} < 1$ & $-3.1\pm0.023$ & $-2.2\pm0.005$ & $2.8\pm0.037$ & $-2.2\pm0.007$ \\ [-0.0ex]
\hline
\end{tabular}
\label{tab 1}
\end{table}

\begin{figure}
\epsscale{1.0} \plotone{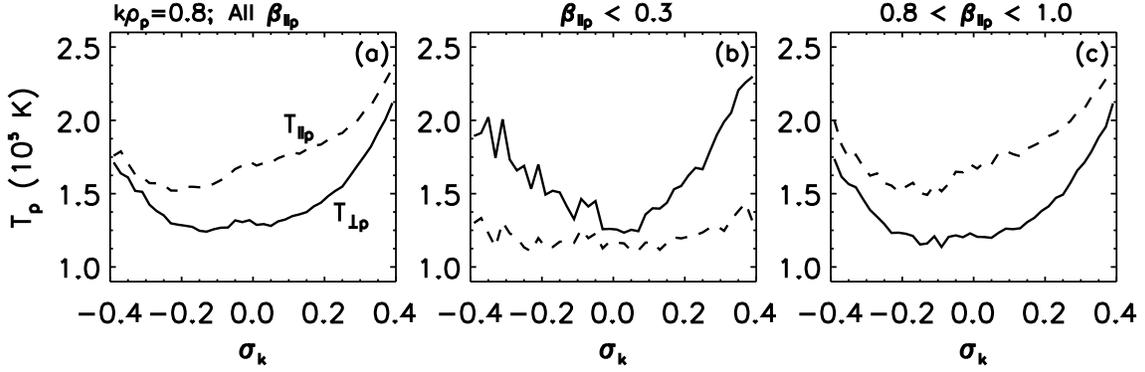} \caption{Medians of proton perpendicular temperature $T_{{\perp}p}$ (solid lines) and parallel temperature $T_{{\parallel}p}$ (dashed lines) against $\sigma_k$ at $k\rho_p=0.8$ for cases of (a) all $\beta_{{\parallel}p}$, (b) $\beta_{{\parallel}p} < 0.3$, (c) $0.8 < \beta_{{\parallel}p} < 1$.}
\end{figure}

\begin{figure}
\epsscale{0.55} \plotone{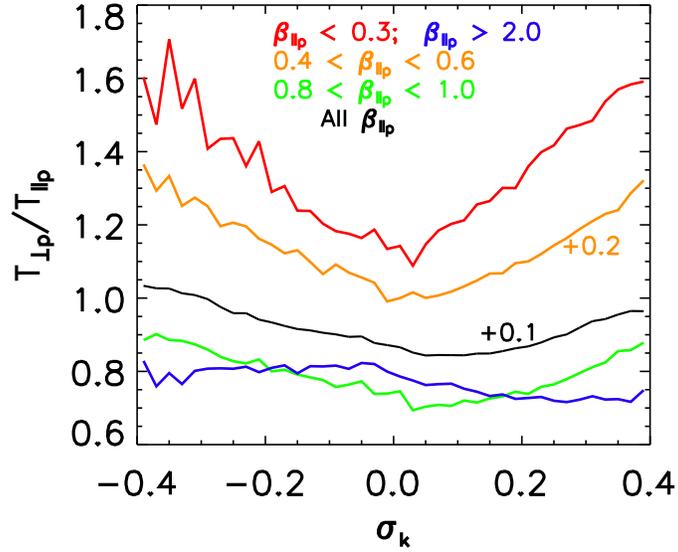} \caption{Medians of proton temperature ratio $T_{{\perp}p}/T_{{\parallel}p}$ against $\sigma_k$ at $k\rho_p=0.8$ for cases of $\beta_{{\parallel}p} < 0.3$ (red line), $0.4 < \beta_{{\parallel}p} < 0.6$ (orange line), $0.8 < \beta_{{\parallel}p} < 1$ (green line), $\beta_{{\parallel}p} >2$ (blue line), and all $\beta_{{\parallel}p}$ (black line). The orange and black lines have been up-shifted by adding 0.2 and 0.1, respectively.}
\end{figure}

\begin{figure}
\epsscale{1.0} \plotone{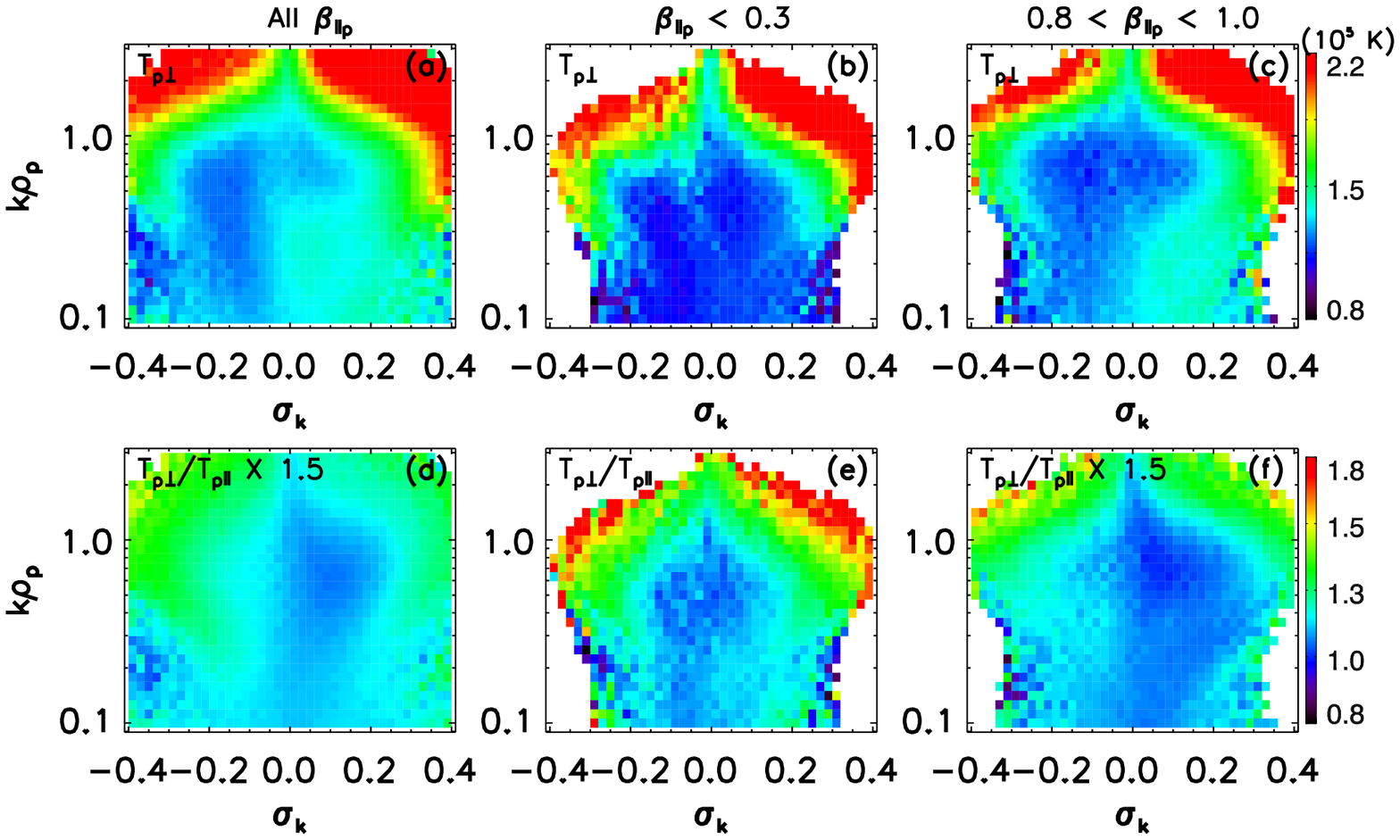} \caption{Color scale plots of medians of $T_{{\perp}p}$ (upper panels) and $T_{{\perp}p}/T_{{\parallel}p}$ (lower panels) in the ($\sigma_k$, $k$) space. Panels (a)$-$(c), as well as panels (d)$-$(f), are for cases of all $\beta_{{\parallel}p}$, $\beta_{{\parallel}p} < 0.3$, and $0.8 < \beta_{{\parallel}p} < 1$, respectively.}
\end{figure}

\subsection{Proton Temperatures Regulated by $\sigma_k$}
This subsection is presented to show how the helicity regulates proton temperatures. Figure 6 plots medians of proton perpendicular and parallel temperatures ($T_{{\perp}p}$ and $T_{{\parallel}p}$) against $\sigma_k$, respectively, with a given wavenumber $k\rho_p=0.8$ as an example. Same as the figures above, three panels in this figure correspond to the three ranges of $\beta_{{\parallel}p}$. From panel (a), one may find the regulations of $\sigma_k$ on the temperatures. They are (1) both $T_{{\perp}p}$ and $T_{{\parallel}p}$ increase with $|\sigma_k|$ when $|\sigma_k|$ is larger than some threshold, 0.15$\sim$0.2, depending on the sign of $\sigma_k$; (2) $T_{{\perp}p}$ tends to increase faster relative to $T_{{\parallel}p}$; (3) the temperature curves are asymmetrical with respect to $\sigma_k=0$ and positive $\sigma_k$ tends to correspond to higher temperatures. According to panel (b), the result appears to be very clear for the low beta case, i.e., $\beta_{{\parallel}p} < 0.3$. In this case $T_{{\perp}p}$ and $T_{{\parallel}p}$ show significantly different dependences on $\sigma_k$. $T_{{\perp}p}$ rapidly rises with $|\sigma_k|$ while $T_{{\parallel}p}$ is nearly irrespective of $|\sigma_k|$, and the temperature curves in this case are less asymmetrical. In contrast, the asymmetry for $T_{{\parallel}p}$ becomes evident when $\beta_{{\parallel}p}$ is large, as shown in panel (c) for the case of $0.8 < \beta_{{\parallel}p} < 1$. Consequently positive $\sigma_k$ results in distinctly higher $T_{{\parallel}p}$, and the minimum of $T_{{\parallel}p}$ happens with $\sigma_k$ around $-0.2$.
In addition, it is also interesting that the fastest increase of $T_{{\perp}p}$ with $|\sigma_k|$ occurs in panel (b) with $\sigma_k>0$, in which the spectral index shows the fastest decrease with $|\sigma_k|$ (panel (b) of Figure 5).

For the sake of discussion, we adopt the following perspectives: (1) the increase of temperatures means the occurrence of heating; (2) the faster increase of $T_{{\perp}p}$ ($T_{{\parallel}p}$) than $T_{{\parallel}p}$ ($T_{{\perp}p}$) implies that the heating occurs preferentially in the direction perpendicular (parallel) to the background magnetic filed. With these perspectives, further investigation shows that the preferentially perpendicular heating occurs when $\beta_{{\parallel}p} \lesssim 1$, while the preferentially parallel heating tends to happen if $\beta_{{\parallel}p} > 2$. To illustrate this point, Figure 7 plots medians of the temperature ratio $T_{{\perp}p}/T_{{\parallel}p}$ against $\sigma_k$ at $k\rho_p=0.8$, where the lines with different colors correspond to different ranges of $\beta_{{\parallel}p}$. To avoid the overlapping of these lines, the orange and black lines have been up-shifted by adding 0.2 and 0.1, respectively. One can see that $T_{{\perp}p}/T_{{\parallel}p}$ almost always rises as $|\sigma_k|$ increases when $\beta_{{\parallel}p} < 1$. On the other hand, $T_{{\perp}p}/T_{{\parallel}p}$ shows somewhat reduction with increasing positive $\sigma_k$ when $\beta_{{\parallel}p} > 2$. Further investigation shows that the dependence of $T_{{\perp}p}$ on $\sigma_k$ becomes much weak if $\beta_{{\parallel}p} > 2$, while $T_{{\parallel}p}$ moderately increases with $\sigma_k$ for $\sigma_k \gtrsim -0.1$ (not shown).

To display the results in Figures 6 and 7 with various wavenumbers, Figure 8 plots medians of $T_{{\perp}p}$ (upper panels) and $T_{{\perp}p}/T_{{\parallel}p}$ (lower panels) against ($\sigma_k$, $k$) for the three cases of $\beta_{{\parallel}p}$ range. To highlight the color comparison in panels (d) and (f), $T_{{\perp}p}/T_{{\parallel}p}$ has been multiplied by 1.5 in both panels. Strong dependences can be found for $0.3 \lesssim k\rho_p \lesssim 1$, where both $T_{{\perp}p}$ and $T_{{\perp}p}/T_{{\parallel}p}$ in principle increase with $|\sigma_k|$. For larger wavenumber with $k\rho_p > 1$, the dependences tend to remain but are not very clear, where $|\sigma_k|$ significantly drops according to Figure 3. Note that $T_{{\perp}p}$ appears to be higher for a larger $k$, which should be attributed to the selection criteria that just allow the time segments with greater turbulent energy to survive (due to the effect of noise at the larger $k$). The higher temperature resulting from the greater magnetic energy at proton scales has been found in the previous research \citep{zha20}. In addition, consistent with the result in Figure 6, the approximate symmetry with respect to $\sigma_k=0$ occurs in the case of $\beta_{{\parallel}p} < 0.3$, i.e., panels (b) and (e), though the larger $\beta_{{\parallel}p}$ tends to break the symmetry according to panels (c) and (f).

\section{Summary and Discussion}
Based on 15 years of in situ measurements, this paper performs a statistical research on the magnetic helicity, magnetic energy spectra, and proton temperatures in the solar wind. Results from the magnetic helicity distributions show that the helicity signature with moderately high magnitude ($0.1 < |\sigma_k| < 0.4$) frequently arises in solar wind turbulence at scales $0.3 \lesssim k\rho_p \lesssim 1$. The distributions are generally asymmetrical, with the helicity mostly positive. There are two peaks in the distributions, occurring at $\sigma_k \simeq \pm0.22$ for $k\rho_p=0.8$. The peaks are weak for the low beta case, i.e., $\beta_{{\parallel}p} < 0.3$, while they can be strong for larger $\beta_{{\parallel}p}$. The magnitude of the helicity depends on $\beta_{{\parallel}p}$ as well as $k$. For a given $k$, the helicity magnitude increases with $\beta_{{\parallel}p}$ when $\beta_{{\parallel}p} < 1$ in principle, but decreases if $\beta_{{\parallel}p} > 1$. This increase with $\beta_{{\parallel}p}$ is consistent with the prediction by hybrid simulations of two-dimensional turbulence with beta less than unity \citep{mar13p62,mar16p015}.

The magnetic helicity appears to play an important role in regulating magnetic spectral indices at proton scales. The spectral indices will become significantly more negative if the helicity magnitudes are larger. This correlation between the spectral indices and helicity magnitudes is particularly clear for the case of $\beta_{{\parallel}p} < 0.3$. By fitting the data at $k\rho_p = 1$, we obtain an expression of the correlation as $\alpha_k=-3.5\sigma_k -2.1$ for positive helicity. In this case the correlation has the steepest slope. It will become slightly flatter if the helicity is negative or $\beta_{{\parallel}p}$ is larger, as shown in Table 1 for details.

The magnetic helicity, on the other hand, also play a considerable role in regulating proton temperatures. Overall, proton temperatures ($T_{{\perp}p}$ and $T_{{\parallel}p}$) usually increase with helicity magnitudes at $0.3 \lesssim k\rho_p \lesssim 1$. The temperature increases show different behaviors in different cases of beta ranges. In the case of $\beta_{{\parallel}p} < 0.3$, it is clear that $T_{{\perp}p}$ fastly increases as $|\sigma_k|$ increases, while this trend is very weak for $T_{{\parallel}p}$. The increase of $T_{{\perp}p}$ faster than $T_{{\parallel}p}$ also occurs in the case of $0.8 < \beta_{{\parallel}p} < 1$. (An opposite result happens if $\beta_{{\parallel}p} > 2$.) The $T_{{\perp}p}$ and $T_{{\parallel}p}$ curves against $\sigma_k$ are asymmetrical, with positive $\sigma_k$ contributing to higher $T_{{\perp}p}$ and $T_{{\parallel}p}$. The asymmetry is more obvious for $T_{{\parallel}p}$ when $\beta_{{\parallel}p}$ is large. The investigation on the temperature ratio $T_{{\perp}p}/T_{{\parallel}p}$ reveals that $T_{{\perp}p}/T_{{\parallel}p}$ almost always increases as $|\sigma_k|$ increases when $\beta_{{\parallel}p} < 1$ (Figure 7), which is consistent with the result of faster increase of $T_{{\perp}p}$ than $T_{{\parallel}p}$ (Figure 6).

The magnetic helicity signature discussed in this paper should mainly result from proton-scale KAW turbulence. A lot of researches on the nature of solar wind turbulence at proton scales support the KAW turbulence model \citep{bal05p02,how08p04,sah09p02,sah10p01,sal12p09,che13p02,gro18p01}.
Our statistical examination in terms of the long-axis direction of magnetic fluctuations at proton scales also favors the model of KAW turbulence \citep{zha20}. Note that KAW turbulence can naturally raise the non-zero magnetic helicity \citep{how10p49,hej12p08,pod13p29}.
Hence it can be expected that the majority of the solar wind turbulence at proton scales is characterized by the considerable magnetic helicity signature, as shown in this paper.

Further, one may conclude that the majority of KAWs in solar wind turbulence appear to be outward propagating with respect to the Sun. Note that the magnetic helicity in this paper is measured in the spacecraft reference frame, whose sign marks the polarization in the spacecraft frame. KAWs are inherently right-handed polarized waves (positive helicity) in the solar wind reference frame \citep{gar86p31,zha14p07}, but they may appear as left-handed polarized waves (negative helicity) in the spacecraft frame when they propagate toward the Sun, in which the large Doppler-shift effect could result in the polarization reversal. Our results in Figures 2 and 3 show that the measured helicity is mostly with the positive sign, implying the dominance of the right-handed polarization in the spacecraft frame. This mostly positive helicity should imply that the KAWs are usually outward propagating (without polarization reversal).

We interpret the elevation of proton temperatures with enhanced magnetic helicity as the occurrence of proton heating in the solar wind. The heating may be attributed to the dissipation of proton-scale KAW turbulence that comes from the fluctuations in the inertial range by turbulent cascade. In this idea, the inertial-range fluctuations would determine the ability of the heating; the inertial-range fluctuations with higher energy would contribute to larger-amplitude KAW turbulence at proton scales, and therefore have greater ability to heat protons. Consequently, higher proton temperatures could be expected if the turbulence amplitude is larger. Existing researches revealed positive correlation between proton temperatures and magnetic fluctuation level in the inertial range \citep{smi06p11,vec18p04}. Our study particularly demonstrated that higher proton temperatures correlate with the larger turbulent amplitudes at proton scales \citep{zha20}. 

Existing literature also documented that higher proton temperatures are associated with steeper proton-scale turbulent spectra based on 33 event study, and concluded ``This suggests that steeper dissipation range spectra imply greater heating rates" \citep{lea98p75}. The present study is in line with this literature. Our results first agree with the finding of higher proton temperatures associated with steeper proton-scale spectra, since we have showed that the magnetic helicity enhancements can simultaneously correlate with higher proton temperatures and steeper spectra at proton scales. We also speculate a specific process as follows. Some dissipation mechanism efficiently operates and quickly removes energy from the proton-scale turbulence, which results in a steeper proton-scale spectrum (with an index less than $-7/3$). The steeper proton-scale spectrum might induce faster energy transfer from the fluctuations in the inertial range by turbulent cascade so that the power-law spectrum at proton scales could be maintained. This speculation is consistent with existing result that steeper spectral forms at proton scales correspond to greater cascade rates in the inertial range \citep{smi06p85}. According to this process, the greater heating rates with steeper proton-scale spectra would be inherently attributed to more efficient dissipation of the turbulence.

The rise of $T_{{\perp}p}/T_{{\parallel}p}$ with $|\sigma_k|$ for $\beta_{{\parallel}p} < 1$ (Figure 7) is due to the faster rise of $T_{{\perp}p}$ than $T_{{\parallel}p}$; both $T_{{\perp}p}$ and $T_{{\parallel}p}$ in principle increase with $|\sigma_k|$. We interpret this phenomenon as the heating that occurs preferentially in the perpendicular direction with respect to the background magnetic field. In the context of KAW turbulence, two mechanisms could contribute to the perpendicular heating, i.e., cyclotron resonance and stochastic heating. Theoretical researches show that cyclotron resonance is possible between KAWs and protons, causing the perpendicular heating of protons \citep{gar04p05,smi12p08,ise19p63}. Simultaneous observations of wave fluctuations and particle kinetics reveal that KAWs seem to be responsible for the anomalous cyclotron resonance of proton beams, causing the perpendicular heating of proton beams \citep{hej15p31}. In the presence of large-amplitude electromagnetic fluctuations at the proton gyroradius scale, the perpendicular heating by turbulent KAWs can also be expected due to stochastic heating \citep{joh01p21,cha10p03,hop18p15}. The occurrence of the stochastic heating in the solar wind has been studied in recent years, and results support the stochastic heating as an effect mechanism to heating the solar wind  \citep{xia13p90,vec17p11,arz19p53,mar19p43,mar20p30}.

Before concluding, we make a final remark as follows. Figures 6 and 8 show that the temperature distributions against the helicity are asymmetrical, especially for the proton parallel temperature. The cause of this asymmetry is not clear. It seems to imply that different helicity signs, and therefore different wave propagation directions with respect to the Sun, correspond to different heating efficiencies. It has been well known that the relative directions between ion cyclotron waves and alpha$-$proton differential flow significantly affect the cyclotron resonance efficiency \citep{pod11p41,zha19p60,zha20p14}. It is unclear whether the propagation directions of KAWs relative to the differential flow, which usually points away from the Sun in the fast solar wind, also affect the (cyclotron/Landau) resonance efficiency and result in the asymmetry of the temperature distributions. Further research on this issue is desirable.

In summary, based on the long period in situ measurements, this paper investigates the magnetic helicity and its role in regulating magnetic energy spectra and proton temperatures in the solar wind. This study should be helpful to discuss the solar wind turbulence and the nonadiabatic behavior of the solar wind. Note that the present discussion is preliminary and further researches are needed.

\acknowledgments
The authors acknowledge the SWE team and MFI team on \emph{Wind} for providing the data, which are available via the Coordinated Data Analysis Web (http://cdaweb.gsfc.nasa.gov/cdaweb/istp$_-$public/). G.Q.Z. is grateful to the hospitality by Auburn University in USA as a visiting scholar. This research was supported by NSFC under grant Nos. 41874204, 41974197, 41674170, 41531071, 11873018, 41804163, 11903016, and supported partly by scientific projects from Henan Province (19HASTIT020,16B140003).


\end{document}